\documentclass[apj]{emulateapj}  
\slugcomment{November 2017} 
\usepackage{dcolumn}
\usepackage{bm}
\usepackage{graphicx}
\usepackage{epsfig}
\usepackage{color}
\usepackage{wasysym}
\usepackage{natbib}
\usepackage{twoopt}
\usepackage{dashrule}

\definecolor{slateblue}{rgb}{0.2,0.2,0.6}
\usepackage[breaklinks=true,colorlinks=true,linkcolor=slateblue,citecolor=slateblue,urlcolor=slateblue,pdfauthor={Tomassetti},pdftitle={TimeLag}]{hyperref}
\def\MyTitle#1{{\section{#1}}}

\newcommand{\ApJ}{ApJ}
\newcommand{\AeA}{A\&A}
\newcommand{\PRL}{PRL}
\newcommand{\PRD}{PRD}
\newcommand{\ASR}{ASR}

\newcommand{\PLB}{PLB}
\newcommand{\MNRAS}{MNRAS}

\newcommand{\AMS}{\textsf{AMS}}
\newcommand{\etal}{et al.}
 
\newcommand{\ie}{\textit{i.e.}} 

\newcommand{\epratio}{{e\ensuremath{^{+}}}/{e\ensuremath{^{-}}}}
\newcommand{\pbarp}{\textsf{\ensuremath{\bar{p}/p}}}
\newcommand{\DT}{\ensuremath{\Delta{T}}}
\newcommand{\Kappa}{\ensuremath{\textbf{K}}}

\newcommand{\p}{\textsf{\ensuremath{p}}}
\newcommand{\pbar}{\textsf{\ensuremath{\bar{p}}}}
\newcommand{\eplus}{\textsf{\ensuremath{e^{+}}}}
\newcommand{\eminus}{\textsf{\ensuremath{e^{-}}}}

\usepackage{verbatim}

\begin{document}
\title{Evidence for a Time Lag in Solar Modulation of Galactic Cosmic Rays}
\shorttitle{Evidence for a Time Lag in Solar Modulation of Galactic Cosmic Rays}
\shortauthors{Nicola Tomassetti \etal}
\author{Nicola Tomassetti$^{\,1}$, Miguel Orcinha$^{\,2}$, Fernando Bar\~{a}o$^{\,2}$, Bruna Bertucci$^{\,1}$}
\address{$^{1}$\,Universit{\`a} degli Studi di Perugia \& INFN-Perugia, I-06100 Perugia, Italy;}
\address{$^{2}$\,Laborat{\'o}rio de Instrumenta\c{c}{\~a}o e F{\'i}sica Experimental de Part{\'i}culas, P-1000 Lisboa, Portugal}
\begin{abstract}
The solar modulation effect of cosmic rays in the heliosphere is an energy-, time-, and particle-dependent phenomenon that
arises from a combination of basic particle transport processes such as diffusion, convection, adiabatic cooling, and drift motion. 
Making use of a large collection of time-resolved cosmic-ray data from recent space missions, we construct a simple predictive model of 
solar modulation that depends on direct solar-physics inputs: the number of solar sunspots and the tilt angle of the heliospheric current sheet. 
Under this framework, we present calculations of cosmic-ray proton spectra, positron/electron and antiproton/proton ratios, 
and their time dependence in connection with the evolving solar activity. 
We report evidence for a time lag $\DT=8.1\pm\,1.2$\,months, between solar-activity data
and cosmic-ray flux measurements in space, which reflects the dynamics of the formation of the modulation region.
This result enables us to \emph{forecast} the cosmic-ray flux near Earth well in advance by monitoring solar activity. 
\end{abstract}
\maketitle

%%%%%%%%%%%%%%%%%%%%%%%%%%%
\MyTitle{Introduction}  %%%
%%%%%%%%%%%%%%%%%%%%%%%%%%%
%
In recent years, new-generation experiments of cosmic-ray (CR) detection have reached an unmatched level of precision
that is bringing transformative advances in astroparticle physics \citep{AmatoBlasi2017,Grenier2015}.
Along with calculations of CR propagation in the galaxy,
the interpretation of the data requires a detailed modeling of the so-called \emph{solar modulation effect}.
Solar modulation is experienced by all CR particles that enter the heliosphere to reach our detectors near Earth.
Inside the heliosphere, CRs travel through a turbulent magnetized plasma, 
the solar wind, which significantly reshapes their energy spectra. This effect is known to change with time,
in connection with the quasi-periodical $11$ year evolution of the 
solar activity and to provoke different effects on CR particles and antiparticles \citep{Potgieter2013,Potgieter2014}. 

Observationally, an inverse relationship between solar activity (often monitored by the number of solar sunspots)
and CR flux intensity
has been known 
about for a long time. The effect of solar modulation in the low-energy CR
spectra ($E\lesssim$\,GeV) is measured by several experiments \citep{Bindi2017,ValdesGaliciaGonzalez2016}.
Theoretically, the paradigm of CR propagation in the heliosphere has been developed soundly over the past decades.
Solar modulation is caused by a combination of basic transport processes such as diffusion, convection, adiabatic cooling, and drift motion,
yet the underlying physical mechanisms and their associated parameters remain under active investigation.
In CR astrophysics, solar modulation is often treated using simplified models where the key parameter(s) are 
degenerated with the parameters of CR propagation in the galaxy.
The development of solar modulation models can be ensured by two crucial factors: 
(i) the precise knowledge of the interstellar spectra (LIS) of CRs outside the heliosphere;
(ii) the availability of time-series of CR data on different species. 
Recent accomplishments from strategic space missions have enabled us to make significant progress in this field. 
The entrance of Voyager-1 in the interstellar space
provided us with the very first LIS data on CR protons and electrons \citep{Stone2013,Cummings2016}. 
Long-duration space experiments PAMELA (on orbit since 2006) and \AMS{} (since 2011)
have started releasing a continuous stream of time-resolved  data on CR particles and antiparticles \citep{Adriani2013,Adriani2016,Bindi2017}. 
These measurements add to a large wealth of low-energy data collected in the last decades by space missions CRIS/ACE \citep{Wiedenbeck2009}
IMP-7/8 \citep{GarciaMunoz1997}, \emph{Ulysses} \citep{Heber2009}, EPHIN/SOHO \citep{Kuhl2016},
and from ground data provided continuously by the neutron monitor (NM) worldwide network \citep{Mavromichalaki2011,Steigies2015}.

In this \emph{Letter}, we present new calculations of CR fluxes near-Earth
that account for the dynamics of CR modulation in the expanding heliosphere.
Using a large collection of modulated and interstellar CR data collected in space,
we have constructed a predictive and measurement-validated model of solar modulation that depends
only on direct solar-activity observables: the sunspot number (SSN) and the tilt angle of the heliospheric current sheet (HCS).
In our calculations, all relevant processes are formally expressed using
\emph{retarded} relations in order to account for a \emph{time-lag}, \DT, 
between solar-activity observations and effective conditions of the modulation region. 
Regarding this region as a bubble with radius $d$$\sim$100-120\,AU and radially flowing
wind of speed $V$$\sim$300-700\,km/s, we expect a time-lag of the order of \DT$\sim$0.5-1\,year.
Such a lag was reported in a number of empirical correlation studies between NM rates and solar-activity
indices \citep{MavromichalakiPetropoulos1984,Badruddin2007,AslamBadruddin2012,AslamBadruddin2015,Belov2005,Lantos2005,Nymmik2000} 
although it is usually ignored in CR modulation models.
As we will show, recent direct measurements of CRs reveal the existence of an eight-month lag.
This result sets the timescale of the changing conditions of the heliosphere, enabling us to
\emph{forecast} the near-Earth CR flux well in advance.
Along with the intrinsic interest in plasma or solar astrophysics, our results address a prerequisite
for modeling space weather effects, which is an increasing concern for space missions and air travelers.
\\

%%%%%%%%%%%%%%%%%%%%%%%%%%
\MyTitle{Methodology}  %%%
%%%%%%%%%%%%%%%%%%%%%%%%%%
%
The transport of CRs in heliosphere is described by the Krymsky-Parker equation for the omnidirectional phase
space density  $\psi(t,p,{\bf  r})$ expressed as a function of time $t$, momentum $p$, and position ${\bf r}$ \citep{Potgieter2013}:
\begin{equation}\label{Eq::TransportEquation}
  \frac{\partial \psi}{\partial t} = - ({\bf V} + {\bf v_{d}}) \cdot \nabla \psi + \nabla \cdot ({\Kappa} \cdot \nabla \psi) + \frac{1}{3}(\nabla \cdot {\bf V}) \frac{\partial \psi}{\partial \ln{p}}
\end{equation}
The various terms represent convection with the solar wind of speed ${\bf V}$, 
drift motion with average speed ${\bf v_{d}}$, spatial diffusion with tensor ${\Kappa}$, and adiabatic momentum losses. % 
We set up a minimal 2D description, with ${\bf r}=(r,\theta)$, radius and helio-colatitude \citep{Bobik2012}. 
The parallel component of the diffusion tensor is $K_{\parallel}=  \kappa^{0} \frac{10^{22}\beta p/{\rm GeV}}{3B/B_{0}}$,
in units of cm$^{2}$/s, where we have factorized an adimensional scaling factor, $\kappa^{0}$, of the order of unity.
The perpendicular component is $K_{\perp}\cong 0.02\,K_{\parallel}$ \citep{GiacaloneJokipii1999}.
The regular magnetic field (HMF) is modeled with the usual Parker structure, with 
$B=\frac{AB_{0}r_{0}^{2}}{r^{2}}\sqrt{1+ \Gamma^{2}}$, with $\Gamma=\frac{\Omega{r}}{V}\sin{\theta}$,
where $\Omega=2.866\cdot\,10^{-6}$\,rad\,s$^{-1}$ is the angular rotation of the Sun,
$B_{0}r^{2}_{0}\cong$\,3.4\,nT\,AU$^{2}$ sets the HMF scale, and $A=\pm\,1$ sets the magnetic polarity cycle of the Sun.
The polarity is positive (negative) when the HMF points outward (inward) in the northern hemisphere.
This model accounts for gradient and curvature drift effects. 
In particular, drift is important across the wavy layer of the HCS, \ie{} the surface where polarity
changes from north to south. The angular extension of the HCS is described by the tilt angle $\alpha$. 
The drift velocity components $v_{r}$ and $v_{\theta}$ are proportional to 
$\frac{2 r \beta p }{q{A}}$, thus the sign of ${\bf {v}_{d}}$ depends on the product $qA$ \citep{BurgerPotgieter1989}. 
We account for the presence of the termination shock (TS), placed at a default distance $r_{\rm ts}\cong$\,85\,AU, 
and for its time dependence over the solar cycle \citep{Manuel2015,VosPotgieter2016}. 
Beyond the TS, plasma density and $B$ increase by a factor $s_{\rm ts}=$\,3, the TS compression ratio, while $V$ and $K$ decrease by $1/s_{\rm ts}$. 
These quantities are changed gradually, between their pre- and post-shock values, across a TS thickness of $L=1.2$\,AU.
For instance, the wind speed is modeled as 
$V(r)=\frac{s_{\rm ts}+1}{2s_{\rm ts}}V_{0} - \frac{s_{\rm ts}-1}{2s_{\rm ts}}V_{0}\tan^{-1}(\frac{r-r_{\rm ts}}{L})$ 
\citep[from Eq.\,3 in][]{Langner2003}, with $V_{0}\cong$\,400\,km\,s$^{-1}$.
A similar procedure, based on an $\arctan$-like function of $\theta$, is also applied to regularize the drift velocity 
near the HCS \citep[from Eq.\,22 and below in][]{Bobik2012}.
The modulation boundaries are set at the heliopause at $r_{\rm hp}=\,122$\,AU. 
The CR distribution in the heliosphere is computed using the stochastic differential equation approach of \citet{Kappl2016}.
This method consists in the backward-in-time propagation of a large number of pseudo-particles from Earth to the boundaries \citep{Raath2016,Strauss2012,AlankoHuotari2007}. 
For a given particle type, steady-state solution of Eq.\,\ref{Eq::TransportEquation} ($\partial\psi/\partial{t}=0$) are obtained by sampling.
The LIS fluxes  $J^{\rm IS}$ are calculated within an improved model of CR acceleration and propagation \citep{Tomassetti2015TwoHalo,TomassettiDonato2015,Feng2016}.
Proton and electron LISs are well constrained by the Voyager-1 and \AMS{} data \citep{Cummings2016,Aguilar2015Proton,Aguilar2015Helium}.
Antiproton and positron LISs rely on secondary production calculations and are affected by larger uncertainties \citep{Feng2016,Tomassetti2012Isotopes}.
The CR number density is given by $N(p)dp=4\pi p^{2}\psi dp$. The resulting flux is $J=\frac{\beta c}{4\pi}N$,
that we express in unit of GeV$^{-1}$\,m$^{-2}$\,s$^{-1}$\,sr$^{-1}$.
%
%%%%%%%%%%%%%%% SOLAR DATA %%%%%%%%%%%%%%%%%%%%
\begin{figure}[!t]
\centering
\includegraphics[width=0.43\textwidth]{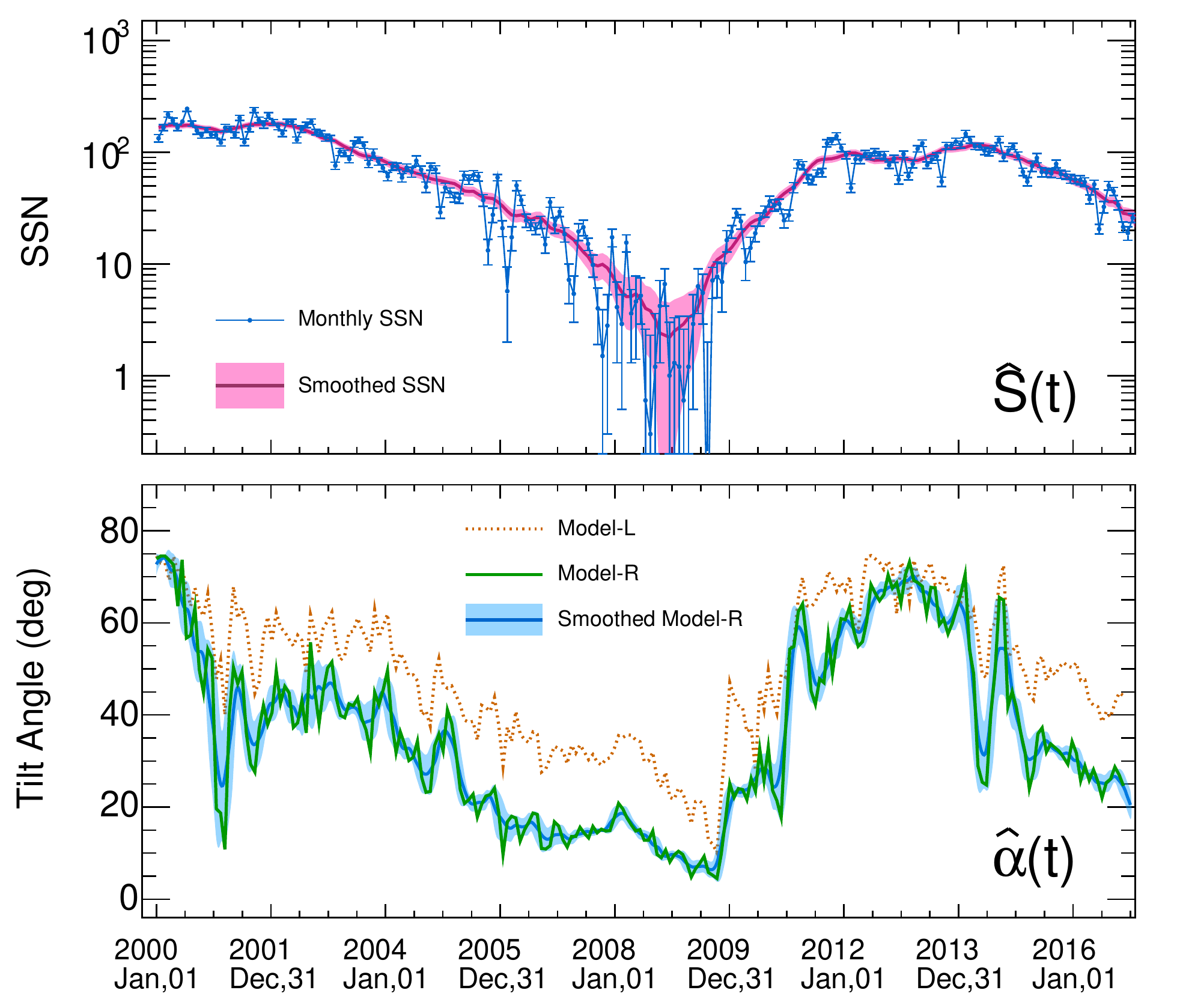} 
\caption{\footnotesize{
    Reconstruction of the sunspot number ($\hat{S}(t)$) and tilt-angle of the heliospheric current sheet ($\hat{\alpha}(t)$) as function of time. 
}}
\label{Fig::ccSolarData}
\end{figure}
%%%%%%%%%%%%%%%%%%%%%%%%%%%%%%%%%%%%%%%%%%%%%%%%%%%%
Due to drift, CR particles and antiparticles sample different parts of the heliosphere.
During negative polarity, protons and positrons ($q>0$) travel near the HCS and drift across that plane,
while electrons and antiprotons ($q<0$) propagate preferentially through the polar regions, 
with faster diffusion and smaller losses. 
The role of positive and negative particles interchanges with polarity.
Along with $qA$, the interplay of the various processes depends on the
levels of HCS waviness and HMF irregularities that are contained in $\alpha$ and $\kappa^{0}$.
Tilt-angle measurements $\hat{\alpha}$ are provided on a 10-day basis from the \emph{Wilcox Solar Observatory} 
using two reconstruction methods: the ``classic'' model-$L$, and the improved model-$R$ \citep{Hoeksema1995,FerreiraPotgieter2004}.
A smoothed interpolation of Model-$R$ is adopted as default.
The diffusion coefficient is also time-dependent, due to changes in the HMF turbulence \citep{Manuel2014}.
A basic diagnostic for the HMF turbulence is the manifestation of solar sunspots \citep{Plunian2009,Boschini2017,Bobik2012}. 
Here, we adopt a simple two-coefficient relation $\kappa^{0} \equiv a+b\log({\hat{S}})$, 
where the $\hat{S}$-function smoothly interpolates the monthly series of SSNs provided by the 
\emph{SIDC - Royal Observatory of Belgium} \citep{CletteLefevre2016}. 
%%%%%%%%%%%%%%% PROTON_TIME_PROFILE %%%%%%%%%%%%%%%%%%%%
\begin{figure*}[!t]
\centering
\includegraphics[width=0.82\textwidth]{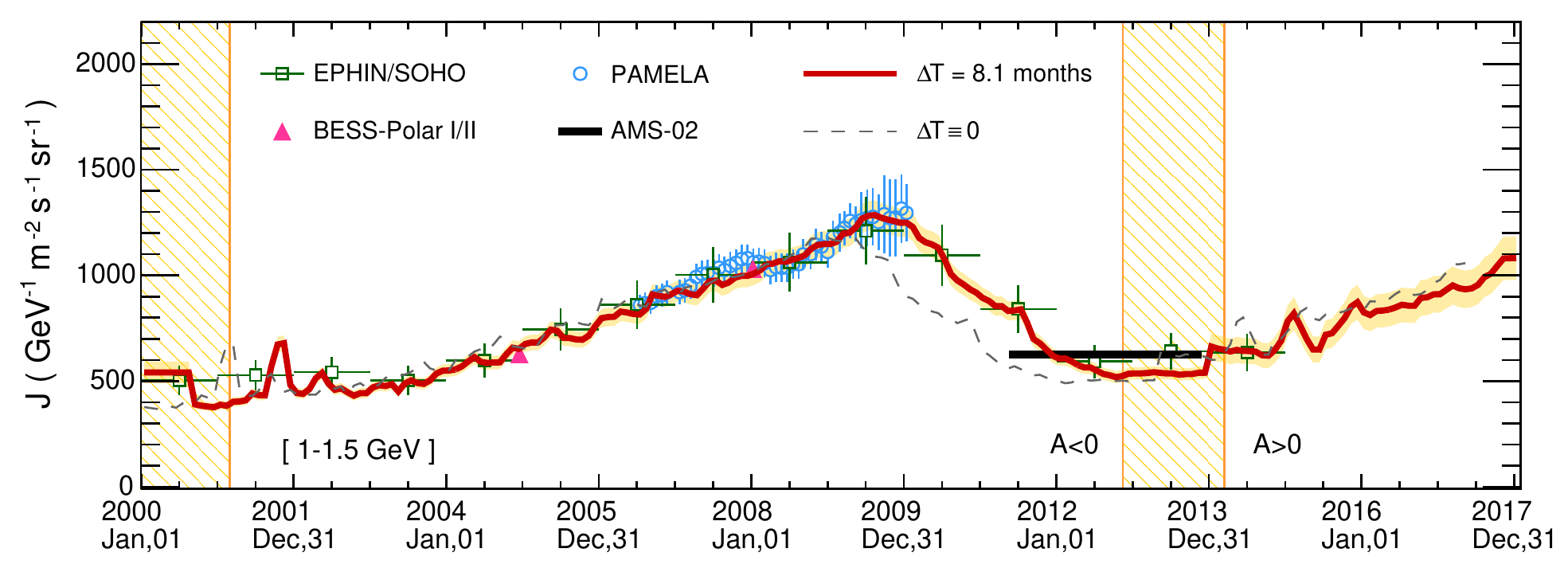}
\caption{\footnotesize{%
Time profile of the proton flux at $E=1-1.5$\,GeV. Best-fit calculations are shown as thick solid line,
along with the uncertainty band, in comparison with the data \citep{Adriani2013,Kuhl2016,Abe2008,Abe2012,Aguilar2015Proton}.
Calculations for $\DT=0$ are shown as thin dashed lines. The shaded bars indicate the magnetic reversals of the Sun's polarity \citep{Sun2015}.
}}
\label{Fig::ccProtonTimeProfile}
\end{figure*}
%%%%%%%%%%%%%%%%%%%%%%%%%%%%%%%%%%%%%%%%%%%%%%%%%%%%
%
%
%%%%%%%%%%%%%%% PROTON_FLUX_2D %%%%%%%%%%%%%%%%%%%%
\begin{figure}[!t]
\centering
\includegraphics[width=0.43\textwidth]{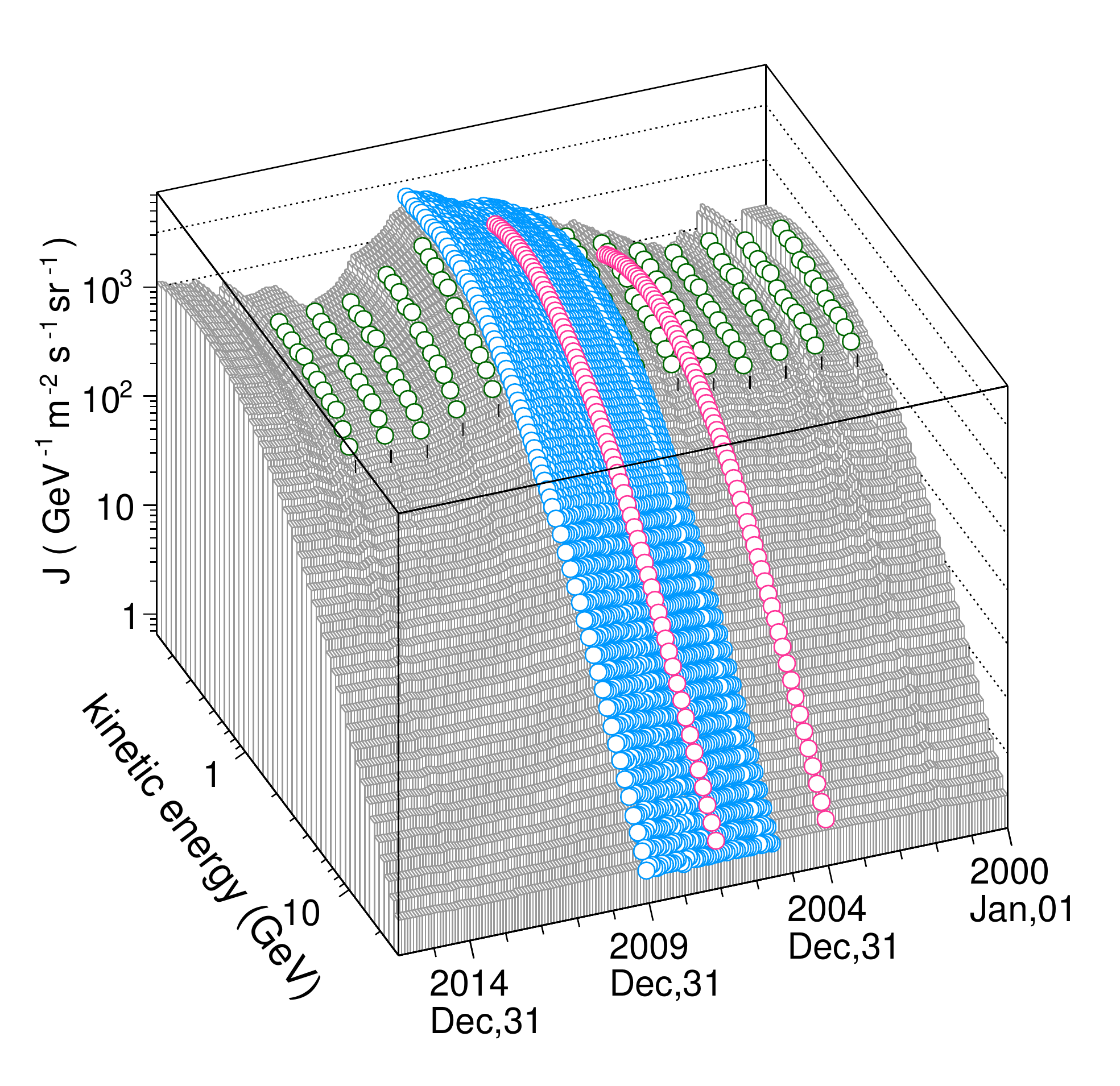}
\caption{\footnotesize{%
Proton flux calculations as function of energy and time in comparison with the
data from PAMELA, EPHIN/SOHO, and BESS \citep{Adriani2013,Kuhl2016,Abe2008,Abe2012}.
}}
\label{Fig::ccProtonFlux2D}
\end{figure}
%%%%%%%%%%%%%%%%%%%%%%%%%%%%%%%%%%%%%%%%%%%%%%%%%%%%
%

The temporal behavior of these quantities, shown in Fig.\,\ref{Fig::ccSolarData}
from 2000 to 2017, is at the basis of the time-dependent nature of solar modulation.
In practice, the problem is modeled under a \emph{quasi-steady} fashion, \ie, by providing a
time-series of steady-state solutions corresponding to a time-series of input parameters.
While this approach is justified by the different timescales of CR transport and solar activity,
the correspondence between solar-activity observations and effective conditions of the heliosphere
may require that dynamics considerations come into play.
A finite amount of time is needed, in fact, for the properties observed in the solar corona
to be transported in the outer heliosphere by the plasma. 
To tackle this issue, we introduce a parameter \DT{} in our calculation, describing
a time-lag between the solar-activity indices of Fig.\,\ref{Fig::ccSolarData} and
the medium properties of the modulation region, \ie, the spatial region effectively sampled by CRs. 
As discussed, evidence for a lag of $\sim$\,6-12 months has been reported in NM-based
empirical studies \citep{MavromichalakiPetropoulos1984,Badruddin2007,Belov2005,Lantos2005,Nymmik2000,MishraMishra2016,Chowdhury2016}.

Our model is specified by three free parameters only, $a$, $b$, and \DT, that we constrain using a large amount of data.
We use monthly-resolved proton data from the PAMELA experiment \citep{Adriani2013}
collected in solar minimum conditions between July 2006 and January 2010, 
and new data from the EPHIN/SOHO space detector \citep{Kuhl2016}, yearly-resolved between 2000 and 2016.
We also include data from the BESS Polar-I (Polar-II) mission from 13 to 21 December 2004 (23 December 2007 to 16 January 2008) \citep{Abe2008,Abe2012}.
These measurements are given in terms of time-series of energy spectra, $\hat{J}_{j,k}=\hat{J}(t_{j},E_{k})$,
each one representing a \emph{snapshot} of the CR flux near-Earth at epoch $t_{j}$.
Calculations $J(t_{j}, E_{k})$  are performed using retarded functions of the physics inputs 
$\alpha_{j}=\hat{\alpha}(t_{j}-\DT)$ and  $\kappa^{0}_{j} = a+b\log(\hat{S}(t_{j}-\DT))$.
We then build a global $\chi^{2}$-estimator:
\begin{equation}\label{Eq::GlobalChiSquare}
\chi^{2}(a,b,\DT) = \sum_{j,\,k} \left[ \frac{ J(t_{j}, E_{k}; a, b, \DT) - \hat{J}_{j,k} }{\sigma_{j,k} } \right]^{2} 
\end{equation}
The quantity $\sigma_{j,k}$ includes experimental errors in the data and model uncertainties due to finite statistics
of the simulation. The following sources of errors are also accounted:
(i) LIS uncertainties from the constraints provided by Voyager-1 and \AMS{} data;
(ii) uncertainties on $\hat{\alpha}(t)$ from the $L$-$R$ model discrepancy, and from
the different smoothing procedures on the HCS drift \citep[see][]{BurgerPotgieter1989}; 
(iii) uncertainties on $\hat{S}(t)$ from the smoothed SSN variance, see Fig.\,\ref{Fig::ccSolarData};
(iv) uncertainties on the TS strength and position including its time variations.
The free parameters are estimated by means of standard minimization techniques.
In practice, for both polarities, calculations are 
performed 
by interpolation over 
4D grids of $\alpha$-values ranging from 0 to 80 degrees with 1-degree step,
$\kappa^{0}$-values ranging from 0.1 to 6 with 0.1 of average (but nonequidistant) resolution step,
$E$-values consisting of 40 log-spaced points between 0.08\,GeV and 50\,GeV, 
and 7 TS position values $r_{\rm ts}$ ranging from 70\,AU to 100\,AU. 
This task required the simulation of about 14 billion proton trajectories, corresponding to several months CPU time.
Along with protons, simulations of \pbar{} and $e^{\pm}$ particles have been carried out.
\\

%%%%%%%%%%%%%%%%%%%%%%
\MyTitle{Results}  %%%
%%%%%%%%%%%%%%%%%%%%%%
%
The global fit has been performed on 3993 proton data points collected between 2000 and 2012 (in $A<0$ conditions)
at kinetic energy between 0.08 and  50\,GeV. The best-fit parameters are
$\hat{a}=$\,4.07\,$\pm$\,0.95, $\hat{b}=$\,-1.39\,$\pm$\,0.34, and $\widehat{\DT}$=8.1\,$\pm$\,1.2\,months, giving $\chi^{2}/df=2643/3990$.
The fit was repeated after fixing $\DT\equiv\,0$, \ie, under the conventional ``unretarded'' scenario,
returning $\hat{a}=$\,3.52\,$\pm$\,0.91, $\hat{b}=$\,-1.16\,$\pm$\,0.28, and $\chi^{2}/df=$\,4979/3991.
Results are shown in Fig.\,\ref{Fig::ccProtonTimeProfile} and Fig.\,\ref{Fig::ccProtonFlux2D},
which illustrate the time and energy dependence of calculations in comparison with the data.
The fits perform well at all energies and epochs.
From Fig.\,\ref{Fig::ccProtonTimeProfile}, it can be seen that the model reproduces 
very well the time evolution of the proton flux, at $E=1.5$\,GeV.
In the $A>0$ period, \ie, 
from the 2013 reversal and up to 2018,
the proton flux is predicted to increase with time. 
The post-reversal time evolution of the proton flux is currently being measured by the \AMS{} experiment \citep{Bindi2017}. 
%
%%%%%%%%%%%%%%% KAPPA vs TIME/SSN %%%%%%%%%%%%%%%%%%%%
\begin{figure}[!t]
\centering
\includegraphics[width=0.43\textwidth]{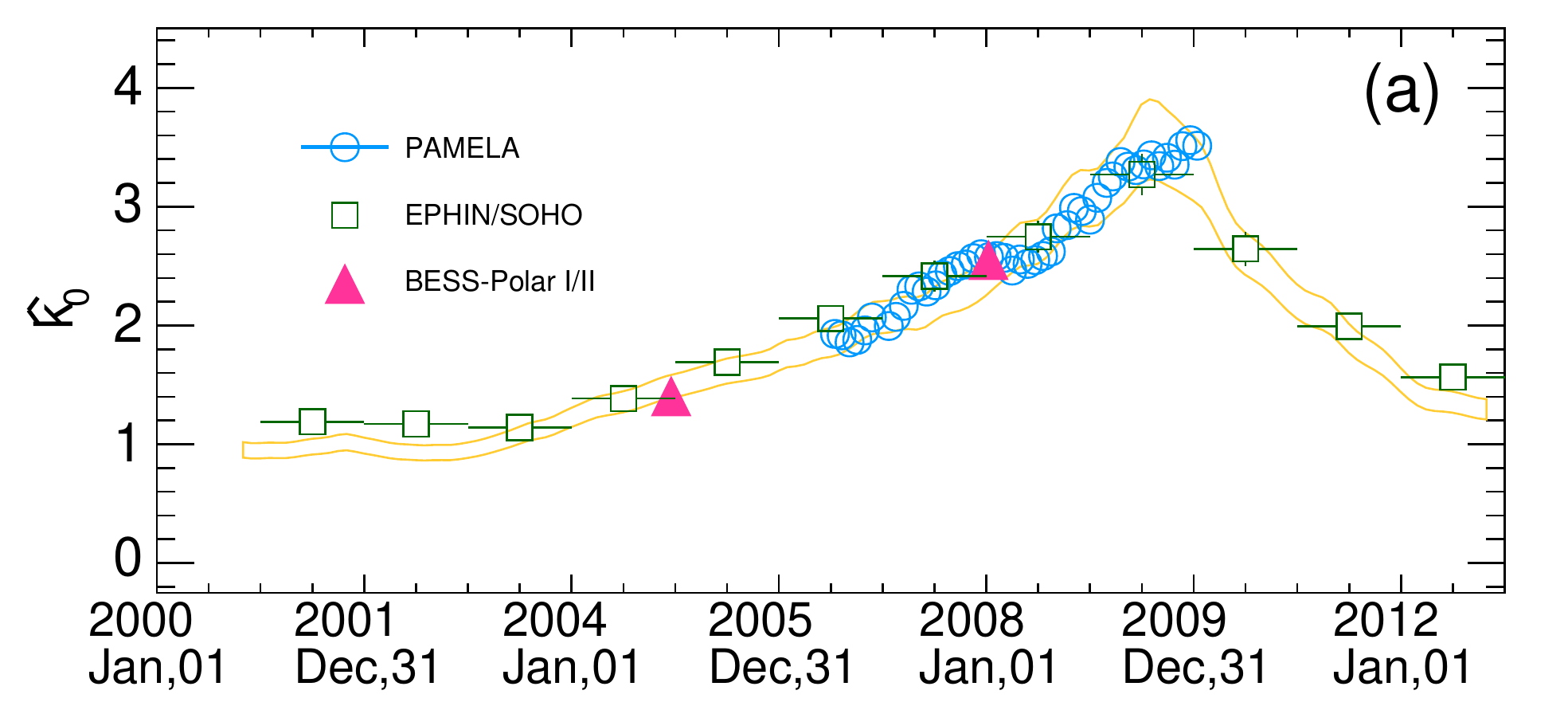}
\includegraphics[width=0.43\textwidth]{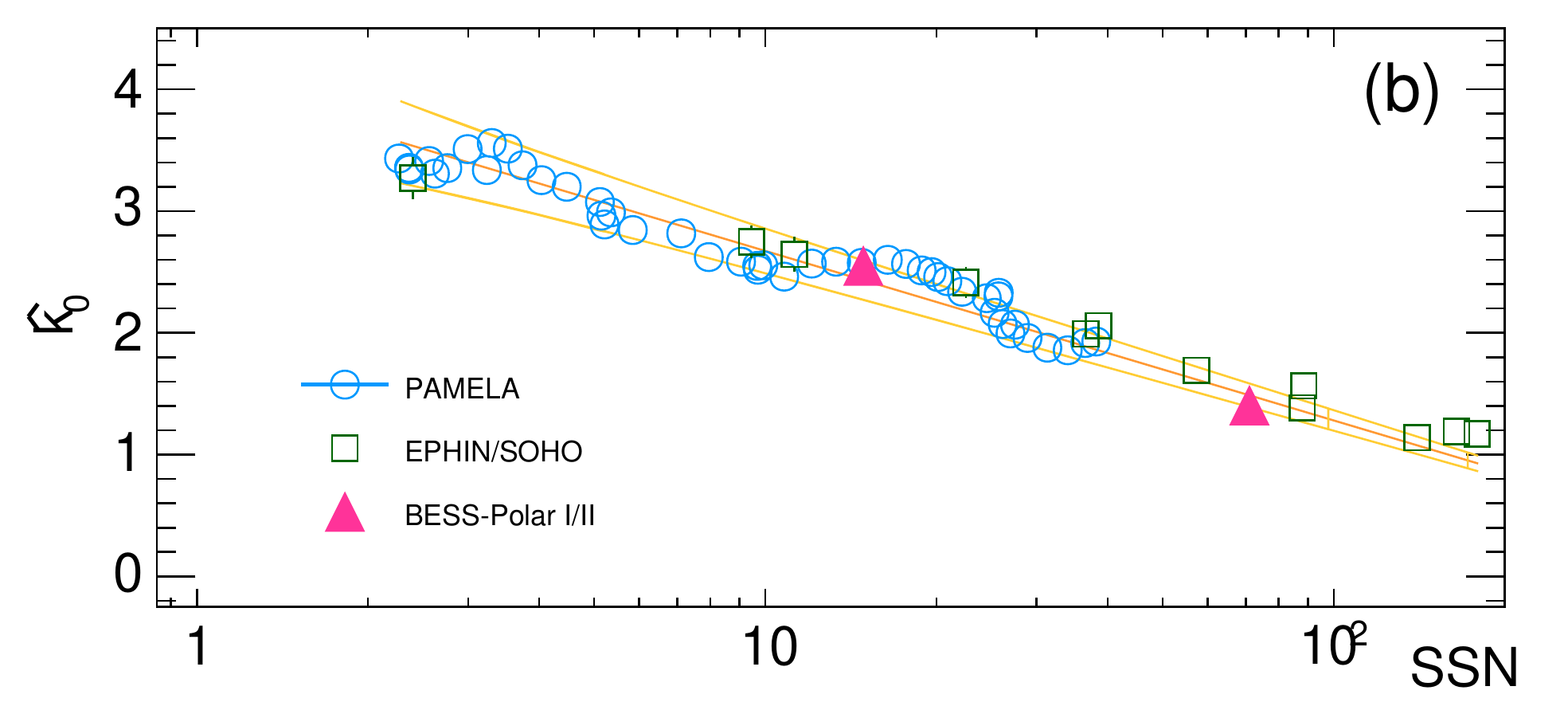}
\includegraphics[width=0.43\textwidth]{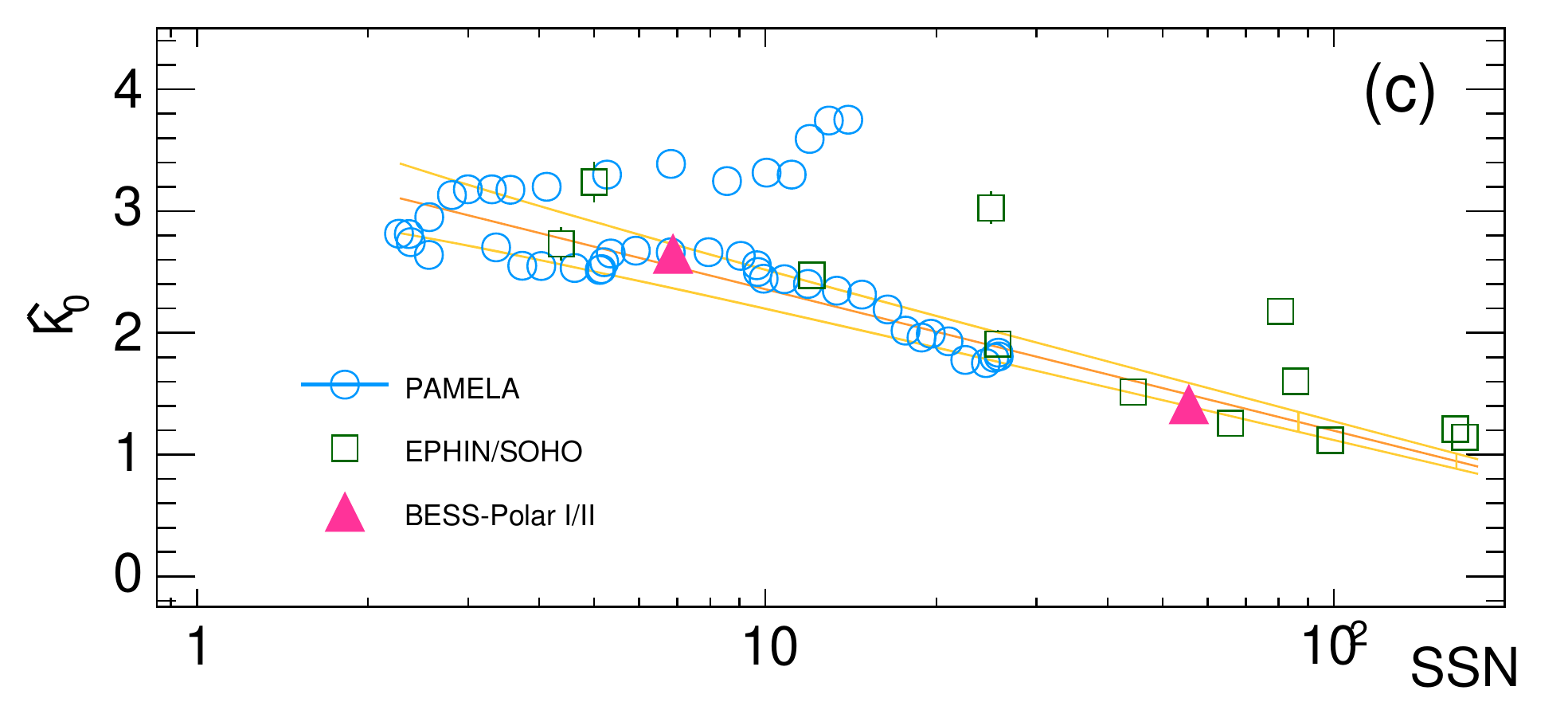}
\caption{\footnotesize{%
Scatter plots representing the best-fit $\hat{\kappa}^{0}$-values
from various data sets \citep{Adriani2013,Kuhl2016,Abe2008,Abe2012}
versus time (a) and versus SSNs after accounting (b) and not accounting (c) for the time lag.
The yellow lines represent the adopted parameterization and its uncertainty band.
}}
\label{Fig::ccKAPPAvsSSN}
\end{figure}
%%%%%%%%%%%%%%%%%%%%%%%%%%%%%%%%%%%%%%%%%%%%%%%%%%%%%%
%
It is also clear that the retarded scenario (with \DT$\equiv$\,8.1\,months, thick red line)
allows for a much better description of the time evolution of the proton flux.
The results presented here are obtained using a \emph{static} TS position at the default value of $r_{\rm ts}=$\,85\,AU.
Extreme values of $r^{\pm}_{\rm ts}=95/75$\,AU have also been tested.
They lead to a $\pm$\,10\,\% shift in the near-Earth flux intensity, 
for fixed parameters, in agreement with \citet{Manuel2015}. 
However, this shift is re-absorbed by different best-fit parameters $\hat{a}$ and $\hat{b}$,
while consistent \DT{} values are inferred in all cases.
Finally, we have tested the case of a \emph{dynamic} TS by means of a time-dependent $r_{\rm ts}(t)$ function.
The TS time dependence is poorly known and must rely on calculations.
We adopt a reference function based on \citet{RichardsonWang2011,RichardsonWang2012} with a rescaling 
in order to match the average TS positions determined 
from calculations based on Voyager-1 and 2 \citep{WebberIntriligator2011,Washimi2011,Provornikova2014}. 
The robustness of the results is studied using a parametric form of the type $c_{1}r_{\rm ts}(t+c_{2})$ with 
$c_{1}=1\pm\,0.12$ and $c_{2}=\pm\,3$\,yrs, along with $s_{\rm ts}=3\pm\,1$. 
We estimated a $\pm$\,0.8\,months of uncertainty for \DT, which is accounted as systematic error in the quoted results.

To inspect our results in greather depth, we have also performed a time-series of fits to single
energy spectra $\hat{J}_{j}$ by directly using the diffusion scaling as a free parameter.
This provided time-series of 62 $\hat{\kappa}^{0}_{j}$-values corresponding to various epochs $t_{j}$.
Fits have been done after fixing $\DT\equiv\,\widehat{\DT}=$\,8.1\,months and $\DT\equiv$\,0, respectively.
The reconstructed time evolution of $\hat{\kappa}^{0}(t)$ is shown in Fig.\,\ref{Fig::ccKAPPAvsSSN}a.
At this point one can inspect the correlation between $\hat{\kappa}^{0}$ and SSN, which is shown
in Fig.\,\ref{Fig::ccKAPPAvsSSN}b and Fig.\,\ref{Fig::ccKAPPAvsSSN}c for the two scenarios, \ie, with and without time-lag.
In the first case, the relation between $\hat{\kappa}^{0}_{j}$ and the ``delayed'' SSN
be well described by a simple universal relation between $\kappa^{0}$ and SSN,
shown as yellow line together with its uncertainty band.
The resulting linear correlation coefficient is $\rho_{\DT}=-0.89$.
In contrast, under the standard \DT=0 scenario, a simple one-to-one correspondence between diffusion
scaling and SSN observations cannot be established along the entire variation range of SSNs.
%
%%%%%%%%%%%%%%% ANTIMATTER/MATTER %%%%%%%%%%%%%%%%%%%%
\begin{figure*}[!t]
\centering
\includegraphics[width=0.82\textwidth]{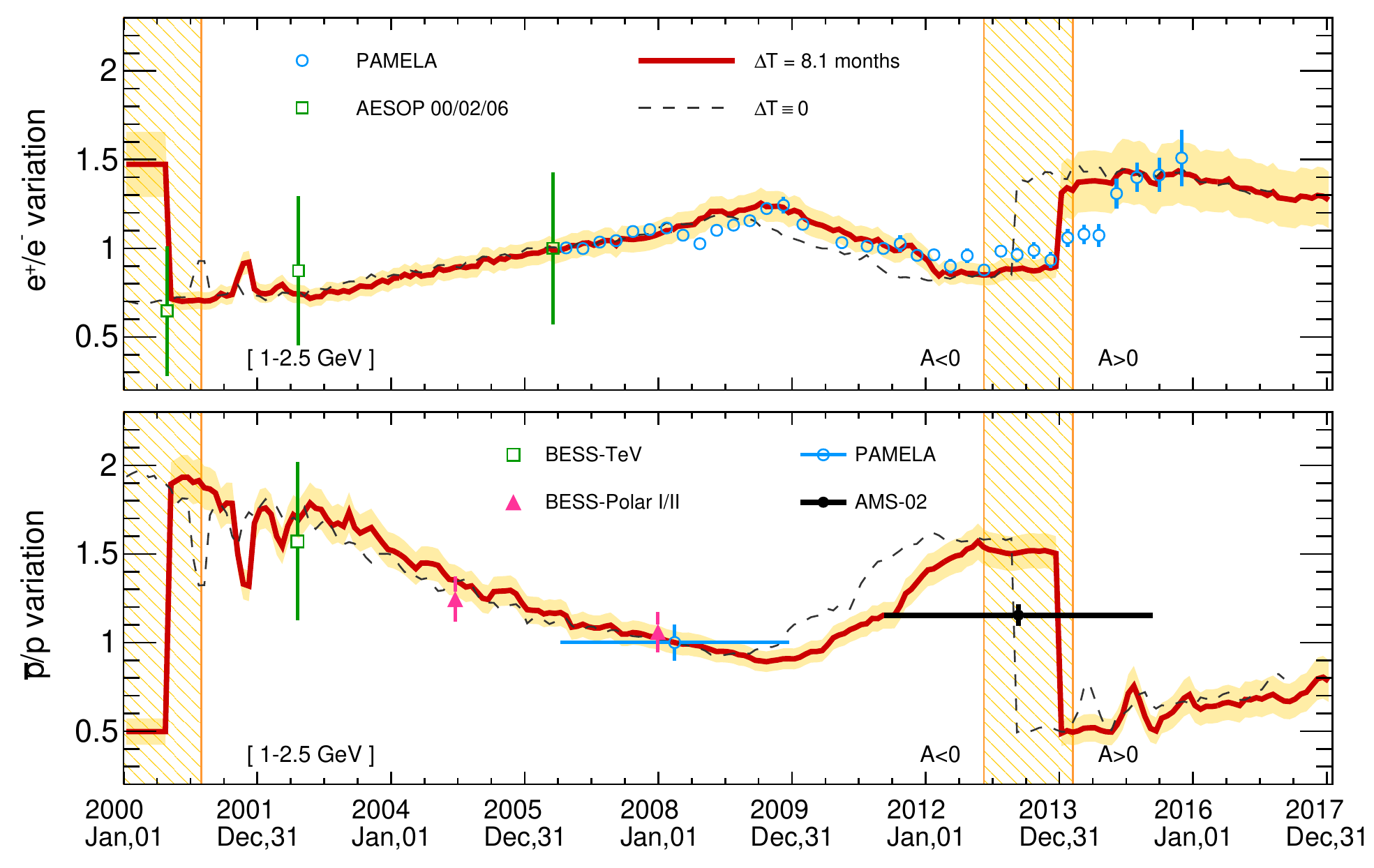}
\caption{\footnotesize{%
Time profile of the ratios \epratio{} (top) and \pbarp{} (bottom) at $E=1\--2.5$\,GeV.
Model predictions and their corresponding uncertainties, are shown in comparisons
with the data \citep{Adriani2016,Abe2008,Abe2012,Clem2009,Haino2005,Aguilar2016PbarP}.
The shaded bars indicate the magnetic reversals of the Sun's polarity \citep{Sun2015}.
}}
\label{Fig::ccAntimatterMatterRatios}
\end{figure*}
%%%%%%%%%%%%%%%%%%%%%%%%%%%%%%%%%%%%%%%%%%%%%%%%%%%%
%
In particular, any $\kappa^{0}(\hat{S})$ relation would fail in describing the transition between 
descending and ascending phases in the solar minimum of 2009
\ie, in proximity to the transition between descending and ascending phases of solar activity.
In this case, the correlation coefficient is $\rho_{0}=-0.64$.
Similar considerations can be drawn from the comparison between the PAMELA data 
in Fig.\,\ref{Fig::ccProtonTimeProfile} and the SSN evolution of Fig.\,\ref{Fig::ccSolarData}b:
the CR proton flux keeps increasing for a few months after solar activity minimum.
These findings explain why other authors, when proposing simple relations between $\kappa^{0}$ and SSN,
had to adopt different coefficients for descending and ascending phases \citep{Bobik2012,Boschini2017}.
Remarkably, this problem is naturally resolved in our model:
once the time-lag is properly accounted for, the $\kappa^{0}$-SSN relation can be captured by a simple universal function.
Similar results are also found from the correlations between $\kappa^{0}$ and the tilt angle.
The two quantities are anticorrelated, with coefficients  $\rho_{\DT}=-0.78$ and $\rho_{0}=-0.57$ corresponding to the two scenarios.
The inferred lag is also found to be highly stable to changes 
in the fitting energy range (moving the minimal energy from $0.08$ to 8\,GeV)
and in the input spectrum (changing the LIS intensity within a factor of two at $E\approx$1\,GeV).

Finally, our model is used to predict the time evolution of antimatter-to-matter ratios such as \epratio{} and \pbarp.
Calculations are shown in Fig.\,\ref{Fig::ccAntimatterMatterRatios} under both scenarios \DT=8.1\,months and \DT=0.
Measurements of the relative variation of the ratios are from PAMELA \citep{Adriani2016}, 
AESOP \citep{Clem2009}, BESS-Polar \citep{Abe2008,Abe2012}, BESS-TeV \citep{Haino2005}, and \AMS{} \citep{Aguilar2016PbarP}.
For the \epratio{} ratio, we note that calculations within $\DT=8.1$\,month are favored, although the data do not permit a resolute discrimination.
Across the magnetic reversal, shown in the figure as shaded bars, a remarkable increase (decrease) of the \epratio{} (\pbarp) ratio is predicted.
It should be noted, however, that the dynamics of the transition cannot be modeled during reversal because the HMF polarity is not well defined.
One may also expect that drift motion is ineffective in this phase.
Nonetheless, a rise in the \epratio{} ratio profile has been detected in the PAMELA data \citep{Adriani2016}, 
and this rise is found to occur a few months after completion of the Sun's polarity reversal.
A corresponding decrease of \pbarp{} ratio is predicted to occur with the same timescale.
Crucial tests can be performed by \AMS{} measurements of the two ratios,
or even better, by measurements of individual particle fluxes for \p, \pbar, \eplus, and \eminus{}
under both polarity conditions and across the reversal.

%%%%%%%%%%%%%%%%%%%%%%%%%%
\MyTitle{Conclusions}  %%%
%%%%%%%%%%%%%%%%%%%%%%%%%%
%
New experiments of CR detection in space have determined the evolution of CR fluxes near Earth with unmatched precision and time resolution.
These data enable us to the investigation of the CR modulation effect and its dynamical connection with the evolving solar activity.
In this  \emph{Letter}, we have reported new calculations of CR modulation based on a physically consistent model
that accounts for particle diffusion, drift, convection and adiabatic cooling.
We have adopted a basic formulation where the time-dependent physics inputs of the model consist in SSN and HCS tilt angle.
We have shown that this model reproduces well the time evolution of the CR proton spectra measured by PAMELA, EPHIN/SOHO, and BESS experiments.
Our model is highly predictive once the correspondence between modulation parameters and solar-activity indices is established.
Our study revealed an interesting aspect of the dynamics of CR modulation in the expanding wind,
that is, the presence of time-lag \DT{} between solar data and the condition of the heliosphere.
Using a large compilation of CR proton data we found $\DT=$\,8.1\,$\pm$\,1.2\,months, which is in agreement with
basic expectations and with recent NM-based analysis \citep{AslamBadruddin2015,MishraMishra2016,Chowdhury2016}.
An interesting consequence of this result is that the galactic CR flux at Earth can be predicted, at any epoch $t$, 
using solar-activity indices observed at the time $t-\DT$. This result is of great interest for real-time
space weather forecast, which is an important concern for human spaceflight.

In this work, the parameter \DT{} has been determined using CR protons during negative polarity.
This parameter has to be viewed as an effective quantity representing
the average of several CR trajectories in the heliosphere during $A<0$ conditions.
Further elaborations may include the use of NM data, for larger observation periods, or the accounting for
a latitudinal dependence in the wind profile or in the diffusion coefficient. 
Since CR particles and antiparticles sample different regions of the heliosphere, we expect slightly different 
time lags, $\DT_{\pm}$ depending on the sign of $qA$ (and in particular, $\DT_{-}\lesssim\,\DT_{+}$).
With the precision of the existing data, we were unable to test this hypothesis. 
A detailed re-analysis of our model, in this direction, will be possible after the release
of monthly-resolved data from \AMS{} on CR particle and antiparticle fluxes.
All calculations presented here will be made available and kept updated at the \textsf{SSDC Cosmic-Ray Database} 
hosted by the \emph{Space Science Data Center} of the \emph{Italian Space Agency} \citep{DiFelice2017}.
\\[0.15cm]
\footnotesize{%
  We thank our colleagues of the AMS Collaboration for valuable discussions.
  %N.T. and B.B. acknowledge the European Commission for support under the H2020-MSCA-IF-2015 action, grant No.\,707543
  %MAtISSE -- \emph{Multichannel Investigation of Solar Modulation Effects in Galactic Cosmic Rays}.
  M.O. acknowledges support from FCT-Portugal, \emph{Funda\c{c}\~{a}o para a Ci\^{e}ncia e a Tecnologia} SFRH/BD/104462/2014.
N.T. and B.B. acknowledge support from MAtISSE - \emph{Multichannel Investigation of Solar Modulation Effects in Galactic Cosmic Rays}.
This project has received funding from the European Union's Horizon 2020 research and innovation programme under the Marie Sklodowska-Curie grant agreement No 707543.

}
\newpage

%%%%%%%%%%%%%%%%%%%%%%%%%%%
  %%
%%%%%%%%%%%%%%%%%%%%%%%%%


\begin{thebibliography}  %%
%%%%%%%%%%%%%%%%%%%%%%%%%%%


%%%%%%%%%%%%%%%%%%%%%%%%%%%%%%%%%%%%%%%%

\bibitem[Abe \etal(2012)]{Abe2012}
Abe, K., Fuke, H., Haino, S., \etal, 
\href{https://dx.doi.org/10.1103/PhysRevLett.108.051102}{\PRL{} 108, 051102 (2012)}

\bibitem[Abe \etal(2008)]{Abe2008}
Abe, K., Fuke, H., Haino, S., \etal,
\href{https://doi.org/10.1016/j.physletb.2008.10.053}{\PLB{} 670, 103 (2008)}

\bibitem[Adriani \etal(2016)]{Adriani2016}
Adriani, O., Barbarino, G. C., Bazilevskaya, G. A., \etal,
\href{http://dx.doi.org/10.1103/PhysRevLett.116.241105}{\PRL{} 116, 241105 (2016)}  

\bibitem[Adriani \etal(2013)]{Adriani2013}
Adriani, O., Barbarino, G. C., Bazilevskaya, G. A., \etal,
\href{http://dx.doi.org/10.1088/0004-637X/765/2/91}{\ApJ{} 765, 91 (2013)}

\bibitem[Aguilar \etal(2016)]{Aguilar2016PbarP}
Aguilar, M., Ali Cavasonza, L., Alpat, B.,
\href{https://dx.doi.org/10.1103/PhysRevLett.117.091103}{\PRL{} 117, 091103 (2016)} 

\bibitem[Aguilar \etal(2015a)]{Aguilar2015Proton}
Aguilar, M., Aisa, D., Alpat, B., \etal, (a), 
\href{http://dx.doi.org/10.1103/PhysRevLett.114.171103}{\PRL{} 114, 171103 (2015)} 
  
\bibitem[Aguilar \etal(2015b)]{Aguilar2015Helium}
Aguilar, M., Aisa, D., Alpat, B., \etal, (b), 
\href{http://dx.doi.org/10.1103/PhysRevLett.115.211101}{\PRL{} 115, 211101 (2015)} 

\bibitem[Alanko-Huotari \etal(2007)]{AlankoHuotari2007}
Alanko-Huotari, K., Usoskin, I. G., Mursula, K., Kovaltsov, G. A.,
\href{https://dx.doi.org/10.1029/2007JA012280}{J. Geophys. Res. 112, A08101 (2007)} 

\bibitem[Amato \& Blasi(2017)]{AmatoBlasi2017}
Amato, E., \& Blasi, P.,
\href{https://doi.org/10.1016/j.asr.2017.04.019}{\ASR, in press (2017)} [\href{https://arxiv.org/abs/1704.05696}{arXiv:1704.05696}]

\bibitem[Aslam \& Badruddin(2012)]{AslamBadruddin2012} 
Aslam, O. P. M., \& Badruddin,
\href{http://dx.doi.org/10.1007/s11207-012-9970-3}{Sol. Phys. 279, 269-288 (2012)}

\bibitem[Aslam \& Badruddin(2015)]{AslamBadruddin2015} 
Aslam, O.P.M. \& Badruddin, 
\href{http://dx.doi.org/10.1007/s11207-015-0753-5}
{Sol. Phys. 290, 2333-2353 (2015)}
  
\bibitem[Badruddin \etal(2007)]{Badruddin2007}
Badruddin, Singh, M., Singh, Y. P.,
\href{https://doi.org/10.1051/0004-6361:20066549}{\AeA{} 466, 697-704 (2007)} 

\bibitem[Belov, \etal(2005)]{Belov2005}
Belov, A. V., Dorman, L.I., Gushchina, R. T., \etal, %, V.N. Obridko, B.D. Shelting, V.G. Yanke
\href{http://dx.doi.org/10.1016/j.asr.2005.03.088}{\ASR{} 35, 491-495 (2005)}

\bibitem[Bindi(2017)]{Bindi2017} 
Bindi, V., Corti, C., Consolandi, C., Hoffman, J., Withman, K.,
\href{https://doi.org/10.1016/j.asr.2017.05.025}{\ASR{} 60, 865-878 (2017)}

\bibitem[Bobik \etal(2012)]{Bobik2012}
Bobik, P., Boella, G., Boschini, M. J., \etal,
\href{https://dx.doi.org/10.1088/0004-637X/745/2/132}{\ApJ{} 745, 132 (2012)}

\bibitem[Boschini \etal(2017)]{Boschini2017}
Boschini, M. J., Della Torre, S., Gervasi, M., La Vacca, G., Rancoita, P. G.,
\href{https://doi.org/10.1016/j.asr.2017.04.017}{\ASR{} in press (2017)}

\bibitem[Burger \& Potgieter(1989)]{BurgerPotgieter1989}
Burger, R. A., \& Potgieter, M. S.,
\href{https://dx.doi.org/10.1086/167313}{\ApJ{} 339, 501-511 (1989)}

\bibitem[Chowdhury \etal(2016)]{Chowdhury2016} 
Chowdhury, P., Kudela, K., Moon, Y.-J.,
\href{https://dx.doi.org/10.1007/s11207-015-0832-7}
{Sol. Phys. 291, 581-602 (2016)} 

\bibitem[Clem \& Evenson(2009)]{Clem2009}
Clem, J., \& Evenson, P.,
\href{https://dx.doi.org/10.1029/2009JA014225}{J. Geophys. Res. 114, A10 (2009)}  

\bibitem[Clette \& Lefevre(2016)]{CletteLefevre2016} 
Clette, F., \& Lef{\`e}vre, L.,
\href{https://dx.doi.org/10.1007/s11207-016-1014-y}
{Sol. Phys. 291, 2629-2651 (2016)}
see also \url{http://www.sidc.be}

\bibitem[Cummings \etal(2016)]{Cummings2016}
Cummings, A. C., Stone, E. C., Heikkila, B. C., \etal,
\href{http://dx.doi.org/10.3847/0004-637X/831/1/18}{\ApJ{} 831, 18 (2016)} 

\bibitem[Di Felice \etal(2017)]{DiFelice2017}
Di Felice, V., Pizzolotto, C., D'Urso, D., \etal, %S. Dari, D. Navarra, R. Primavera, B. Bertucci
\href{https://pos.sissa.it/301/1073/}
{Proc. 35$^{\rm th}$ ICRC - Bexco, PoS 1073 (2017)} %Busan, Korea
see also \url{https://tools.asdc.asi.it/CosmicRays}

\bibitem[Feng \etal(2016)]{Feng2016}
Feng, J., Tomassetti, N., Oliva, A.,
\href{http://dx.doi.org/10.1103/PhysRevD.94.123007}{\PRD{} 94, 123007 (2016)}

\bibitem[Ferreira \& Potgieter(2004)]{FerreiraPotgieter2004}
Ferreira, S. E. S., \& Potgieter, M. S.,
\href{http://dx.doi.org/10.1086/381649/}{\ApJ{} 603, 744 (2004)}

\bibitem[Garcia-Munoz \etal(1997)]{GarciaMunoz1997}
Garcia-Munoz, M., Simpson, J. A., Guzik, T. G., Wefel, J. P., Margolis, S. H.,
\href{http://dx.doi.org/10.1086/191197}{\ApJ\,SS 64, 269-304 (1987)} 

\bibitem[Giacalone \& Jokipii(1999)]{GiacaloneJokipii1999}
Giacalone, J., \& Jokipii, J. R.,
\href{http://dx.doi.org/10.1086/307452}{\ApJ{} 520, 204 (1999)}

\bibitem[Grenier \etal(2015)]{Grenier2015} 
Grenier, I. A., Black, J. A., Strong, A. W.,
\href{https://dx.doi.org/10.1146/annurev-astro-082214-122457}{Annu. Rev. Astron. Astrophys., 53, 199–246 (2015)}

\bibitem[Haino \etal(2005)]{Haino2005}
Haino, S., Abe, K., Fuke, H., \etal,
\href{http://adsabs.harvard.edu/abs/2005ICRC....3...13H}{Proc. 29$^{\rm th}$ ICRC - Pune 3, 13 (2005)}

\bibitem[Heber \etal(2009)]{Heber2009}
Heber, B., Kopp., A., Gieseler, J., \etal,
\href{http://dx.doi.org/10.1088/0004-637X/699/2/1956}{\ApJ{} 699, 1956-1963 (2009)} 

\bibitem[Hoeksema(1995)]{Hoeksema1995} 
Hoeksema, J. T.,
\href{http://dx.doi.org/10.1007/BF00768770}{Space Sci. Rev. 72, 137-148 (1995)}
see also \url{http://wso.stanford.edu}
 
\bibitem[Kappl \etal(2016)]{Kappl2016}
Kappl, R.,
\href{http://dx.doi.org/10.1016/j.cpc.2016.05.025}{Comp. Phys. Comm. 207, 386-399 (2016)} 

\bibitem[K{\"u}hl \etal(2016)]{Kuhl2016}
K{\"u}hl, P., G{\'o}mez-Herrero, R., Heber, B.,
\href{http://dx.doi.org/10.1007/s11207-016-0879-0}{Sol. Phys. 291, 965-974 (2016)}

\bibitem[Langner \etal(2003)]{Langner2003}
Langner, U. W., Potgieter, M. S., Webber, W. R.,
\href{http://dx.doi.org/10.1029/2003JA009934}{J. Geophys. Res. 108, A10, 8039 (2003)}

\bibitem[Lantos(2005)]{Lantos2005}
Lantos, P.,
\href{http://dx.doi.org/10.1007/s11207-005-5565-6}{Sol. Phys. 229, 373 (2005)} 

\bibitem[Manuel \etal(2015)]{Manuel2015}
Manuel, R., Ferreira, S. E. S., Potgieter, M. S.,
\href{https://dx.doi.org/10.1088/0004-637X/799/2/223}
{\ApJ{} 799, 223 (2015)}

\bibitem[Manuel \etal(2014)]{Manuel2014}
Manuel, R., Ferreira, S. E. S., Potgieter, M. S.,
\href{https://dx.doi.org/10.1007/s11207-013-0445-y}
{Sol. Phys. 289, 2207-2231 (2014)}

\bibitem[Mavromichalaki \& Petropoulos(1984)]{MavromichalakiPetropoulos1984}
Mavromichalaki, H. \& Petropoulos, B.,
\href{http://dx.doi.org/10.1007/BF00653915}{Astrophys. Space Sci. 106, 61-71 (1984)} 

\bibitem[Mavromichalaki \etal(2011)]{Mavromichalaki2011} 
Mavromichalaki, H., Papaioannou, A., Plainaki, C, \etal,
\href{http://dx.doi.org/10.1016/j.asr.2010.02.019}{\ASR{} 47 2210-2222 (2011)}

\bibitem[Mishra \& Mishra(2016)]{MishraMishra2016} 
Mishra, V. K. \& Mishra, A. P.,
\href{https://dx.doi.org/10.1007/s12648-016-0895-9}{Indian J. Phys. 90, 1333-1339 (2016)}

\bibitem[Nymmik(2000)]{Nymmik2000}
Nymmik, R. A.,
\href{https://doi.org/10.1016/S0273-1177(99)01242-9}{\ASR{} 26, 1875-1878 (2000)}

\bibitem[Plunian \etal(2009)]{Plunian2009}
Plunian, F., Sarson, G. R., Stepanov, R.,
\href{https://doi.org/10.1111/j.1745-3933.2009.00760.x}{\MNRAS{} 400, L47-L51 (2009)}

\bibitem[Potgieter(2013)]{Potgieter2013}
Potgieter, M. S.,
\href{https://dx.doi.org/10.12942/lrsp-2013-3}{Living Rev. Sol. Phys., 10, 3 (2013)}

\bibitem[Potgieter(2014)]{Potgieter2014}
Potgieter, M. S.,
\href{https://doi.org/10.1016/j.asr.2013.04.015}{\ASR{} 53, 1415–1425 (2014)}

\bibitem[Provornikova \etal(2014)]{Provornikova2014}
Provornikova, E., Opher, M., Izmodenov, V. V., Richardson, J. D., Toth, G.,
\href{https://dx.doi.org/10.1088/0004-637X/794/1/29}
{\ApJ{} 794, 29 (2014)}

\bibitem[Raath \etal(2016)]{Raath2016}
Raath, J. L., Potgieter, M. S., Strauss, R. S., Kopp, A.,
\href{https://doi.org/10.1016/j.asr.2016.01.017}{\ASR{} 57, 1965–1977 (2016)}

\bibitem[Richardson \& Wang(2012)]{RichardsonWang2012}
Richardson, J. D., \& Wang, C.,
\href{https://dx.doi.org/10.1088/2041-8205/759/1/L19}
{\ApJ{} 759, L19 (2012)}

\bibitem[Richardson \& Wang(2011)]{RichardsonWang2011}
Richardson, J. D., \& Wang, C.,
\href{https://dx.doi.org/10.1088/2041-8205/734/1/L21}
{\ApJ{} 734, L21 (2011)}

\bibitem[Steigies(2015)]{Steigies2015}
Steigies, C., %[\url{http://www.nmdb.eu}] % 
\href{http://pos.sissa.it/cgi-bin/reader/contribution.cgi?id=236/225}{Proc. 34$^{\rm th}$ ICRC - The Hague, PoS 225 (2015)} 

\bibitem[Stone \etal(2013)]{Stone2013}
Stone, E. S., Cummings, A. C., McDonald, \etal, %F. B., Heikkila, B. C., Lal, N., and Webber, W. R.,
\href{http://dx.doi.org/10.1126/science.1236408}{Science 341, 6142, 150 (2013)}  

\bibitem[Strauss \etal(2012)]{Strauss2012}
Strauss, R. D., Potgieter, M. S., B{\"u}shing, I., Kopp, A., 
\href{http://dx.doi.org/10.1007/s10509-012-1003-z}{Astrophys. Space Sci. 339, 223 (2012)}

\bibitem[Sun \etal(2015)]{Sun2015}
Sun, X., Hoeksema, J. T., Liu, Y., Zhao, J.,
\href{https://dx.doi.org/10.1088/0004-637X/798/2/114}{\ApJ{} 798, 114 (2015)}

\bibitem[Tomassetti(2015)]{Tomassetti2015TwoHalo}
Tomassetti, N.,
\href{http://dx.doi.org/10.1103/PhysRevD.92.081301}{\PRD{} 92, 081301 (2015)} 

\bibitem[Tomassetti(2012)]{Tomassetti2012Isotopes}
Tomassetti, N.,
\href{http://dx.doi.org/10.1007/s10509-012-1138-y}
{Astrophys. Space Sci. 342, 131-136 (2012)}

\bibitem[Tomassetti \& Donato(2015)]{TomassettiDonato2015}
Tomassetti, N., \& Donato, F.,
\href{http://dx.doi.org/10.1088/2041-8205/803/2/L15}{\ApJ{} 803, L15 (2015)}

\bibitem[Vald{\'e}s-Galicia \& Gonz{\'a}lez(2016)]{ValdesGaliciaGonzalez2016}
Vald{\'e}s-Galicia, J. F., \& Gonz{\'a}lez, L. X.,
\href{https://doi.org/10.1016/j.asr.2015.11.009}{\ASR{} 57, 1294-1306 (2016)}

\bibitem[Vos \& Potgieter(2016)]{VosPotgieter2016}
Vos, E. E., \& Potgieter, M. S.,
\href{https://dx.doi.org/10.1007/s11207-016-0945-7}
{Sol. Phys. 291, 2181-2195 (2016)}


\bibitem[Washimi \etal(2011)]{Washimi2011}
Washimi, H., Zank, G. P., Hu, Q., \etal, %Tanaka, T., Munakata, K.,  Shinagawa, H.,
\href{https://dx.doi.org/10.1111/j.1365-2966.2011.19144.x}
{MNRAS 416, 1475-1485 (2011)}

\bibitem[Webber \& Intriligator(2011)]{WebberIntriligator2011}
Webber, W. R., \& Intriligator, D. S.,
\href{https://dx.doi.org/10.1029/2011JA016478}
{J. Geophys. Res. 116, A06105 (2011)}

\bibitem[Wiedenbeck \etal(2009)]{Wiedenbeck2009}
Wiedenbeck, M. E. Davis, A. J. Cohen, C. M. S., \etal, %0545
\href{http://icrc2009.uni.lodz.pl/proc/pdf/icrc0545.pdf}{Proc. 31$^{\rm st}$ ICRC - Lodz, 0545 (2009)}  


%%%%%%%%%%%%%%%%%%%%%%%%%
\end{thebibliography}
\end{document}